\documentstyle[epsfig,graphicx,11pt,epsf]{article}

\newcommand{\nn}{\nonumber}

\newcommand{\be}{\begin{equation}}
\newcommand{\ee}{\end{equation}}
\newcommand{\ba}{\begin{eqnarray}}
\newcommand{\ea}{\end{eqnarray}}

\newcommand{\Dslash}{{D\hspace{-8pt}/}}
\newcommand{\delslash}{\partial \hspace{-6pt}/}

\newcommand{\eps}[1]{\epsilon_{#1}}

\newcommand{\Nmass}{M_{N}} 
\newcommand{\delmass}{M_{\Delta}} 
\newcommand{\rhomass}{m_\rho} 
\newcommand{\rhocoup}{g_\rho} 
\newcommand{\f}{f} 
\newcommand{\nucfld}{\psi_N} 
\newcommand{\fpiNN}{f_{\pi N N}} 
\newcommand{\fpiND}{f_{\pi N \Delta}} 
\newcommand{\GMquark}{G^M_{(q)}} 

\newcommand{\vecpi}{\vec \pi}
\newcommand{\vectau}{\vec \tau}
\newcommand{\vecrho}{\vec \rho}
\newcommand{\delmu}{\partial_\mu}
\newcommand{\delMu}{\partial^\mu}

\begin{document}
\begin{center}
{\Large{\bf The $\rho NN$ coupling with direct coupling and loops
  }}

\vspace{0.3cm}

\end{center}

\vspace{1cm}

\begin{center}
{\large{D. Jido, E. Oset and J. E. Palomar }}
\end{center}

\begin{center}
{\small{\it Departamento de F\'{\i}sica Te\'orica and IFIC, \\
Centro Mixto Universidad de Valencia-CSIC, \\
Ap. Correos 22085, E-46071 Valencia, Spain}}

\end{center}

\vspace{1cm}

\begin{abstract}
Starting from a gauge formalism of $\rho$ mesons, pions and baryons we evaluate
the $\rho$ coupling to the nucleon, including the direct coupling provided by
the Lagrangians, plus contributions from loops with the virtual pion cloud. We
find a contribution to the magnetic $\rho$ coupling to the nucleon from pionic
loops of the same size as the direct coupling, which is, however, still small
compared to the empirical values. This finding goes in line with chiral
formulations of the strong interaction of mesons at low energies where, unlike
the scalar mesons which are mostly built of a pion (kaon) cloud, the $\rho$
meson stands as a genuine QCD state with intrinsic properties not tied to those of
the pion cloud.
\end{abstract}

\section{Introduction}
The $\rho$ coupling to the nucleon has been the subject of permanent
attention from different points of  view.  The $\rho$ exchange plays an
important role in the nucleon-nucleon interaction at intermediate
energies \cite{Machleidt:1987hj} and the strength of the tensor interaction is
particularly large, about twice the value given by the vector meson
dominance hypothesis (VMD).  The determination of this coupling was done
in \cite{hoeler} using dispersion relations and has been reconfirmed with
posterior analysis along the same lines \cite{Mergell:1996bf}.  The strength 
of the tensor coupling found in \cite{hoeler,Mergell:1996bf} finds also support
 from the
values of the mixing parameter, $\epsilon$, at energies of the nucleons
around or bigger than 200 MeV \cite{Brown:1994pq}.
Attempts to describe this deviation from VMD have been done from
different perspectives.  In \cite{Brown:1986gu} an analysis using a topological
chiral model concluded that the strong tensor coupling could be
described  semiquantitatively in a two phase model with half and half
fractioning of charge and baryon number between the core and the soliton
cloud.  A color dielectric model was used in \cite{Ren:1990fx} concluding
that the ratio $\kappa_\rho / \kappa_v$ was bigger than one.  QCD sum
rules have also been used in \cite{Wen:1997rf}, and more recently in
\cite{Zhu:1999yg} where the tensor coupling is also found big  and compatible
with empirical determinations.  Relativistic
quark models were used  to obtain also the $\rho NN $ coupling in \cite{weber}.
The chiral quark models \cite{Theberge:1980ye,Brown:1980yu} were also used to 
obtain the
vector and tensor coupling in \cite{Ferchlander:1982cm,osetrho}, assuming that
they come solely from the coupling of the $\rho$ to the pion cloud of the
nucleon and determining the radius of the bag to reproduce the empirical
results. In one way or another all these works come to confirm the
important role of the pion cloud in these couplings.

However, stimulated by the interest in determining the renormalization
of the $\rho$ properties in the nuclear medium, much work has been done
recently \cite{Herrmann:1993za,Klingl:1997kf,Urban:1998eg,Urban:2000im,
Cabrera:2000dx} which make a
revision of the problem timely. There has also been progress in another front
 concerning the $\rho$ meson.
Indeed, Chiral Perturbation Theory, ($\chi PT$), as an effective theory of the
underlying QCD, has emerged as a useful tool to deal with hadronic interactions 
at low and intermediate energies. For the meson meson interaction the basic
dynamics is contained in the lowest and second order Lagrangians of Gasser and
Leutwyler \cite{Gasser:1984pr,Gasser:1984ux}. One important step forward in the understanding
 of the content of  these Lagrangians was done in 
\cite{Ecker:1989te} where it was found that the parameters of the second order
Lagrangian could be generated by the explicit exchange of resonances, 
particularly the vector mesons.  Another step forward was subsequently given in 
\cite{Oller:1999zr} where the lowest order chiral Lagrangian, together with 
the exchange of vector and scalar mesons suggested in
\cite{Ecker:1989te}, were used to study the meson meson interaction,
 implementing  unitarity via the N/D method. The result of
this work was that, while the coupling of the vector mesons to the pseudoscalar
mesons was essential to reproduce the data, the coupling of the scalars was
compatible with zero. Yet, the scalar mesons were generated within the unitary
approach simply from the multiple scattering of the mesons driven by the
interaction accounted for by the lowest order Lagrangian.  These mesons are
hence dynamically generated in this approach, contrary to the vector mesons
which qualify as genuine mesons.  This classification can also be linked to
arguments of the large $N_c$ limit in QCD. Indeed, loops are subleading in the
$N_c$ counting, and consequently in the large $N_c$ limit the scalar mesons
would disappear while the formerly called genuine mesons would survive, hence
giving extra meaning to the concept of genuine and dynamically generated 
mesons \cite{Oller:1999zr,Gasser:1990bv,Meissner:1991kz,Oller:1999hw}.

   This distinction is hence more than semantics and has practical repercussions.
Indeed, since the lowest of the scalar mesons, the $\sigma$, is built up here 
from the meson meson interaction, consistency with this picture demands that the
coupling of a $\sigma$ to the nucleon is simply done by coupling the
interacting meson pair to the nucleon and this is what was done in 
\cite{Oset:2000gn}. However, since the $\rho$ meson qualifies as one of the
genuine mesons, its coupling to the nucleon does not have to come from just the
pion cloud within this approach and, indeed, the effective Lagrangians used in 
\cite{Herrmann:1993za,Klingl:1997kf,Urban:1998eg,Urban:2000im,Cabrera:2000dx}
contain a direct coupling of the $\rho$ to the nucleon.

  With this picture in mind, and within the
philosophy of these effective Lagrangians, it is still proper to ask which is the 
contribution to the $\rho NN$ coupling from the mesonic loops in the 
perturbative expansion.

  There is also another new element in the present evaluation since the findings
of the lattest references
\cite{Herrmann:1993za,Klingl:1997kf,Urban:1998eg,Urban:2000im,Cabrera:2000dx} 
 have also shown the
importance of vertex corrections, linked to the underlying gauge
structure of the Lagrangians, which were overlooked in the previous
determinations of the meson cloud contributions to the $\rho$
couplings.  All these findings introduce new elements in the evaluation
of the vector and tensor couplings of the $\rho$  to the nucleon and the
purpose of the present work is to give a new look to the problem from
this modern perspective.

\section{Model for the $\rho NN$ coupling }
We shall calculate the nucleon
coupling to the $\rho$ meson based on the chiral Lagrangian for the
pion and nucleon. The basic couplings of the pion and nucleon to the
$\rho$ meson are introduced by imposing the gauge theory for vector
meson on the chiral Lagrangian.  In the gauge theory for vector meson,
this particle is considered as a gauge boson of an implicit gauge
symmetry, which was originally suggested by Sakurai \cite{Sakurai:1960ju}
with the vector meson dominance hypothesis, where it mediates all hadronic
interactions. Later on, Bando {\it
et al.} \cite{Bando:1985ej} developed this idea to the hidden local
gauge theory of the non-linear sigma model. As we shall see, at
tree level, the gauge condition of the vector meson on the chiral
Lagrangian gives only the vector coupling of the $\rho NN$ vertex,
since the tensor coupling itself is free from the gauge constraint. 
Therefore, in absence of direct tensor contributions, the tensor coupling is
generated through pion-loop
contributions. The study of such contributions is the aim of this paper.

Let us start by considering the elementary $\rho NN$ vertex. According to the
 construction of the  gauge theory, the direct
coupling of a hadron $h$  to the $\rho$ meson is constructed replacing
the derivative by the associated  covariant derivative with the
gauge symmetry :  
\begin{equation}
  \delmu h(x) \rightarrow 
   D_\mu h(x) = \delmu h(x) + i \rhocoup [ h(x) , V^a ] \rho^a_\mu
   \label{rep} 
\end{equation}
where $V^a$ is the generator of the gauge symmetry of the $\rm SU(2)$
isospin space and the hadronic
charge $\rhocoup$ is given as a universal constant in all hadrons.

For the nucleon, since under the isospin rotation it is
transformed as the fundamental representation: $[V^a,\nucfld] = -
{1\over 2} \tau^a \nucfld$, the covariant derivative for the nucleon
is written as
\begin{equation}
    D_\mu \nucfld = (\delmu + i {\rhocoup \over 2} \vectau \cdot
    \vec\rho_\mu ) \nucfld  
\end{equation}
where $\tau^a$ is the Pauli matrix for the isospin.
Thus, the replacement on the kinetic term of the nucleon,
$\bar \nucfld i\delslash \nucfld \rightarrow \bar \nucfld i\Dslash
\nucfld$, gives the direct coupling of the $\rho$ meson to the nucleon:
\begin{equation}
   {\cal L}_{\rho NN} = - {\rhocoup \over 2} \bar \nucfld \gamma^\mu
   \vec\rho_\mu \cdot \vectau \nucfld \label{rhoNN}
\end{equation}
 
\begin{figure}
\begin{center}
  \epsfbox{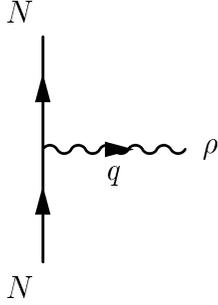}
\end{center}
  \caption{The direct $\rho NN$ coupling. \label{direct_rhoNN}}
\end{figure}

The Lagrangian (\ref{rhoNN}) contributes the vertex shown in
fig.\ref{direct_rhoNN}  as
\begin{equation}
   {-i t}_{\rho NN}^{\rm direct} = - i{\rhocoup \over 2} \tau^a \gamma
   \cdot \varepsilon^*\ ,  \label{tree}
\end{equation}
where $\varepsilon_\mu$ is the polarization vector of $\rho$, and $a$ is the
isospin index of the $\rho$ meson. 
In general, the vertex function of the $\rho NN$ coupling is written
in terms of two Lorentz independent functions $G^V$ and $G^T$:
\begin{equation}
-i t_{\rho NN} = i\left( G^V(q^2) \gamma^\mu + {G^T(q^2)
 \over 2i\Nmass } \sigma^{\mu\nu}q_\nu \right) \varepsilon_\mu^{*}  \tau^a
 \ , \label{GVGT}
\end{equation}
where $q$ is the outgoing momentum of the $\rho$. From eq. (\ref{tree}) the
vertex functions $G^V$ and $G^T$ at tree level are obtained with the result  
\begin{equation}
   G^V_{\rm tree} = - {\rhocoup \over 2}\ , \hspace{0.5cm} G^T_{\rm tree}
   = 0
\end{equation} 

\noindent while empirically one has
\begin{equation}
   G^V_{\rm emp.}= 2.9\pm 0.3 , \hspace{0.5cm} G^T_{\rm emp.} = 18\pm 2
\end{equation} 

In the Lagrangian language, the tensor part is written as
\begin{equation}
   {\cal L}^{\rm tensor}_{\rho NN} = -  { G^T \over 2 \Nmass}\bar \nucfld
   F^{\mu\nu}_{a} \sigma_{\mu\nu} {\tau^a \over 2} \nucfld \label{GTlag}
\end{equation}
with the field strength tensor of the rho meson $F_{\mu\nu}^a = \delmu
\rho_\nu^a - \partial_\nu \rho_\mu^a - \rhocoup \eps{abc} \rho_\mu^b
\rho_\nu^c$.  The Lagrangian of eq.~(\ref{GTlag}) itself is invariant under
the gauge transformation. Therefore the value of the coefficient $G^T$
is free from the constraint of the gauge theory. Here we would like to
calculate $G^T$ from pion-loop contributions as the pion cloud without
introducing any direct tensor couplings. 

For later convenience we work with
the Breit frame, that is $q^0=0$, $p_i=\vec q /2$ and $p_f=-\vec q/2$,
and also use the non-relativistic form. Then eq. (\ref{GVGT}) is
written as 
\begin{equation}
-i t_{\rho NN} =  \left( i G^E( q) \varepsilon^0 -
 {G^M(q) \over 2\Nmass } (\vec \sigma \times \vec q)\cdot \vec
 \varepsilon  \right) \tau^a \ , \label{GEGM}
\end{equation}
with 
\begin{eqnarray}
   G^E(q) &=& G^V(q) \\
   G^M(q) &=& G^T(q) + G^V(q) \ .
\end{eqnarray}

In order to include the contribution from the pion cloud, we calculate the
indirect coupling of the $\rho$ meson to the nucleon shown in
fig.\ref{loop} a).
Since this diagram is one-loop, we also consider other one-loop
diagrams shown in fig.\ref{loop} for consistency of the loop
expansion. As intermediate baryons in the loops, which can be excited
by the pion, we consider both nucleon and $\Delta$. The loop
corrections do not contribute to $G^E$ at $q=0$ due to the
Ward-Takahashi identity.

The $\rho \pi \pi $ coupling is introduced by the gause theory for
vector meson in the same way as the nucleon case. With the
isospin rotation for pions $[V^a, \pi^b] = -i \epsilon^{abc} \pi^c$,
we obtain the covariant derivative for the pion: 
\begin{equation}
   D_\mu \pi^a = \delmu \pi^a + \rhocoup \eps{abc} \pi^b \rho^c_\mu
   \label{pirep} 
\end{equation}
and, then, from the replaced kinetic term ${1 \over 2} D_\mu \vecpi
\cdot D^\mu \vecpi$, we obtain the $\rho \pi\pi$ vertex :
\begin{equation}
  {\cal L}_{\pi\pi\rho} = -\rhocoup (\vecpi \times \delMu \vecpi)
  \cdot \vecrho_\mu
\end{equation}
With this Lagrangian we can calculate the decay width of the $\rho$ meson
decaying to two pions, which gives $\rhocoup = - 6.14$. The sign is given by
comparing to the standard $\rho$ coupling to pions in the chiral
tensor formalism \cite{Ecker:1989te}, which provides the equivalence
$\rhocoup = - \rhomass G_V / \f^2$, where $G_V$ is the parameter appearing in
the chiral resonance Lagrangians of ref.~\cite{Ecker:1989te} providing the $\rho
\pi \pi$ coupling.

For the $\pi N$ couplings, we use the chiral Lagrangian:
\begin{equation}
  {\cal L}_{\pi N} = -{g_A \over 2 \f}\bar{N}\delslash \gamma_5
  \vecpi \cdot \vectau N - {1 \over 4 \f^2} \bar \nucfld (\vecpi
  \times \delslash \vecpi)\cdot \vectau \nucfld+ \cdots \ , \label{piN}
\end{equation}
where $\cdots$ denotes terms with multiple pions which do not
enter the present calculation. In eq.~(\ref{piN}) $g_A$ is the 
axial charge of the nucleon, $g_A = 1.26$. In alternative formulations $g_A$
appears as $D+F$, with $D$ and $F$ the two $SU(3)$ coefficients for the
semileptonic decay of hyperons, or through $g_A /2f\equiv f_{\pi
NN}/m_{\pi}$. The gauge theory for vector
meson introduces the $\pi \rho NN$ coupling through the replacement
of the derivative by the covariant derivative (\ref{pirep}):   
\begin{equation}
  {\cal L}_{\pi\rho N} = { g_A \rhocoup\over 2 \f} \bar{N}
  (\vec\rho_\mu \times \vecpi)\cdot \vectau \gamma^\mu \gamma_5 N +
  {\rhocoup \over 4 \f^2} \bar \nucfld \gamma^\mu (\vecpi\cdot\vecpi
  \vec\rho_\mu \cdot \vectau - \vecpi\cdot\vec\rho_\mu
  \vecpi\cdot\vectau) \nucfld + \cdots \label{pirhoN}
\end{equation}
The first terms of eqs.(\ref{piN}) and (\ref{pirhoN}) give the $\pi NN$
and $\pi \rho NN$ vertices. After the 
non-relativistic reduction we have 
\begin{eqnarray}
   -i t_{\pi NN} &=& -{g_A \over 2 \f} \vec q\cdot \vec \sigma\
    \tau^a  \label{piNN}\\
   -i t_{\pi \rho NN} &=& i {g_A \rhocoup \over 2 \f} \vec\sigma
    \cdot \vec\varepsilon \ \eps{abc} \tau^c \label{pirhoNN}
\end{eqnarray}
where $a$ and $b$ are the isospin indices for $\pi$ and $\rho$,
respectively, and $\vec q$ is the outgoing pion momentum.

For the $\Delta$ contribution, the introduction of $\pi N \Delta$
coupling is empirically performed through the replacement of
the spin-isospin matrix $\vec\sigma$, $\vectau$ on the $\pi NN$ vertex
by the spin-isospin transition matrices $\vec S$, $\vec T$:
\begin{equation}
   -i t_{\pi N\Delta} = -\left({\fpiND \over \fpiNN}\right) {g_A
   \over 2 \f} \vec q\cdot \vec S\ T^a 
\end{equation}
where $(\fpiND/ \fpiNN)=2.13$. If we recall the introduction of the
rho meson coupling through the replacement of eq. (\ref{pirep}), the $\pi N
\Delta$ coupling relates to the $\pi \rho N \Delta$ one. Therefore, in
 analogy with the introduction of the $\pi \rho NN$ coupling
(\ref{pirhoNN}) from the $\pi NN$ vertex (\ref{piNN}), we have
\begin{equation}
   -i t_{\pi \rho N\Delta} = i\left({\fpiND \over \fpiNN}\right){g_A
   \rhocoup \over 2 \f} \vec S \cdot \vec\varepsilon \ \eps{abc} T^c
\end{equation}

For the $\rho \Delta \Delta$  coupling, we use the
gauge theory for vector mesons again. The covariant derivative for the
delta baryon is given in the relativistic form with the
Rarita-Schwinger field for the spin $3/2$ fermion as
\begin{equation}
   D_\mu \psi_{\Delta,\nu} = \delmu \psi_{\Delta,\nu} + i \rhocoup
   \vec T_\Delta \cdot \vec \rho_\mu \psi_{\Delta,\nu}
\end{equation}
where $T_\Delta$ is the isospin $3/2$ matrix, which
is normalized so that $T^3_{\Delta 11} = 3/2$. The kinetic term for the $\Delta$
with the covariant derivative,
which is written as $\bar\psi_{\Delta,\mu} i \Dslash
\psi^\mu_\Delta$, gives the direct coupling of the $\Delta$ to the $\rho$
meson. Similarly to the nucleon case, there exists a tensor
coupling free from the gauge constraint. 
After including the tensor coupling and the non-relativistic
reduction, the $\rho\Delta\Delta$ vertex is written as 
\begin{equation}
   -i t_{\rho\Delta\Delta} = \left( - i \rhocoup \varepsilon^0 - {
    G_{\Delta}^M \over 2 \Nmass } ( \vec S_\Delta \times
    \vec q) \cdot \vec{\varepsilon} \right) T_\Delta^a \label{rhoDD}
\end{equation}
where $S_\Delta$ is the same matrix as the isospin matrix $T_\Delta$
but for spin space, and $G^M_{\Delta}$ is a free parameter. 
Here we assume that the magnetic coupling of the direct
$\rho \Delta \Delta$ is scaled to that of the
$\rho NN$ according to the quark model \cite{Brown:1975di}, which is
\begin{equation}
   G^M_{\Delta} = {4 \over 5 } G^M_{N}
\end{equation}
The derivation is written in Appendix \ref{QM}. If the direct tensor term is
not included, the magnetic coupling comes from the vector term and we obtain
$G^M_{\Delta}= {4 \over 3} {\Nmass \over \delmass} G^M_N$.

For the $\rho N \Delta$ coupling,  the vector coupling is
not allowed due to the spin symmetry, thus, only the magnetic coupling
is allowed and given as
\begin{equation}
   -i t_{\rho N\Delta} = \left( - {
    G_{N\Delta}^M \over 2 \Nmass } ( \vec S \times \vec q) \cdot
    \vec{\varepsilon} \right) T^a \label{rhoDN}
\end{equation}
with the quark model result
\begin{equation}
   G^M_{N \Delta} = {6 \over 5 } \sqrt{2} G^M_{N} 
\end{equation}

\section{\Large{\bf One loop contributions}}

\begin{figure}[ht]
\centering
\epsfysize=6.2cm
\epsfbox{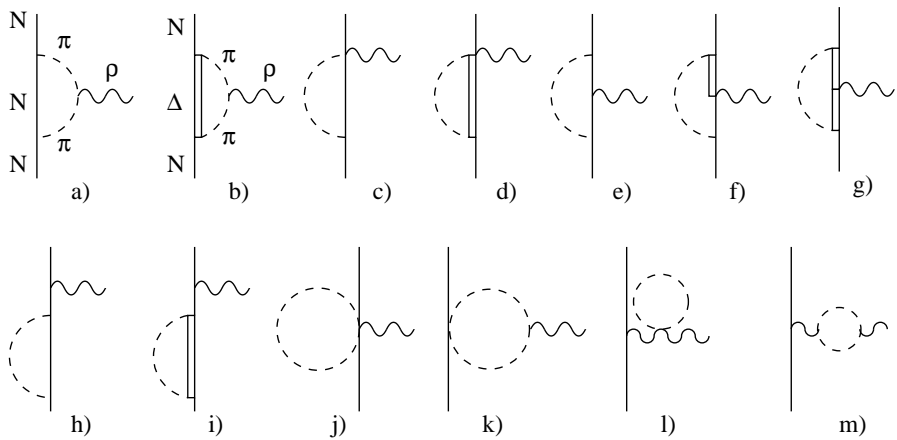}
\caption{One loop diagrams evaluated.}
  \label{loop}
\end{figure}

The evaluation of the tree level diagram with a contact $\rho NN$ interaction
from the Lagrangian of eq. (\ref{rhoNN}) contributes to the $G^{E}_{\rho NN}$ term,
providing a value $G^{E}_{\rho
NN}=\frac{M_{\rho}G_{V}}{2f^{2}}F_{\rho}(\vec{Q}=\vec{0})$=3.07, which is
already a very good value when compared to the experimental one, 
$G^{E}_{\rho NN}=2.9\pm 0.3$ ($F_{\rho}(\vec{Q})$ is the $\rho$ form factor,
defined in Appendix B). However, this is the only contribution to 
$G^{M}_{\rho NN}$ , which has an empirical value around 21. We want to see how much
of the magnetic strength can be generated through loop contributions. In this section we
perform the one loop calculation, which comes from the diagrams shown in figure
 \ref{loop} plus other time orderings. In all of them we assume the external 
 nucleon lines to be protons.

The contribution of each of these diagrams to $G^{E}_{\rho NN}$ and to 
$G^{M}_{\rho NN}$ is given in Appendix B. We discuss here in more detail the
 calculations and the results obtained.
In what follows we present results calculated with the form factors for $\pi$ 
and $\rho$ given in eq.~(\ref{defi}) and we take  $g_A=1.26$ and $f=93$ MeV.

In diagram a) the pions are a $\pi^{+} \pi^{-}$ 
pair  since the $\pi^{0} \pi^{0}$ pair does not couple to the $\rho^{0}$, being
 the intermediate nucleon a neutron. The evaluation of this diagram is 
 straightforward and gives the values $G^{E}_{\rho NN}(\vec{q}=\vec{0})=1.64$
  and $G^{M}_{\rho NN}(\vec{q}=\vec{0})=4.42$. The calculation of diagram b)
  is
analogous, once  the spin and isospin factors arising from eq.
(\ref{repchange}) of Appendix B are taken into account. It gives $G^{E}_{\rho NN}(\vec{q}=\vec{0})
=-1.11$ and $G^{M}_{\rho NN}(\vec{q}=\vec{0})=1.44$. As we can see, the
contributions of these two diagrams to $G^{E}_{\rho NN}$ have opposite sign,
and therefore there is an important cancellation between them, while the contributions to
$G^{M}_{\rho NN}$ have the same sign.

The other diagrams to be considered, except for diagram k), contain vertices in
which the $\rho$ is coupled directly to a nucleon leg. In all these cases we
have multiplied the result provided by the expressions given in Appendix B by
the corresponding $F_{\rho}(\vec{Q})$ form factor, defined in eq.~(\ref{defi}). The 
results given in 
Appendix B for diagram c)  accounts for the diagram
represented in c) of figure \ref{loop} but also for the one with the $\rho$
meson attached to the lower vertex instead of the upper one. In these vertices
the pions are charged. Their contribution to $G^{E}_{\rho NN}$ is of order
${\cal O}(1/M)$ and is not considered here. The contribution to $G^{M}_{\rho
NN}$ is found to be  $G^{M}_{\rho NN}(\vec{q}=\vec{0})=-1.90$. In the case of
 diagram d) there are two
more diagrams contributing since we can have $p\pi^{-}\Delta^{++}$, $\rho
p\pi^{-}\Delta^{++}$ vertices. We obtain from these diagrams 
$G^{M}_{\rho NN}(\vec{q}=\vec{0})=-0.55$, being the contribution to
$G^{E}_{\rho NN}$ of ${\cal O}(1/M)$ as in the previous case. 

In diagrams e), f) and g) we have the $\rho NN$, $\rho N\Delta$ and $\rho \Delta
\Delta$ vertices. The contributions to $G^E$ and $G^{M}$ of these diagrams are
written in Appendix B, where we have taken into account the quark model based
relations between $G^{M}$, $G^{M}_{\Delta}$ and $G^{M}_{N\Delta}$ of Appendix A.
Diagram e) accounts actually for two diagrams, one with an intermediate
$p$ and another one with an intermediate $n$. The evaluation of these
diagrams is simple and gives $G^{E}_{\rho NN}(\vec{q}=\vec{0})=-0.41$ and 
$G^{M}_{\rho NN}(\vec{q}=\vec{0})=0.14$. In the case of diagrams f) and g)
 we have to take into account that they correspond to more diagrams  than in the
case of diagram e) since we can have an intermediate $\Delta^{++}$ in addition
to $\Delta^{+}$ and $\Delta^{0}$. In the evaluation of diagram g) 
we need to apply the 
following relation:

\begin{center}
\be
S_{i}S_{\Delta j}S^{\dagger}_{k}=\frac{5}{6}\textrm{i}\epsilon_{ijk}-
\frac{1}{6}\delta_{ij}\sigma_{k}+\frac{2}{3}\delta_{ik}\sigma_{j}- \frac{1}{6}
\delta_{jk}\sigma_{i}
\label{releses}
\ee
\end{center}

\noindent With this we get $G^{E}_{\rho NN}(\vec{q}=\vec{0})=2.77$ and 
$G^{M}_{\rho NN}(\vec{q}=\vec{0})=0.93$ from these diagrams. We find
also that diagrams of type f) do not contribute to $G^{E}_{\rho NN}$, and they
 provide a contribution to $G^{M}$ of $G^{M}_{\rho NN}(\vec{q}=\vec{0})=1.40$. 

The next diagrams that we have considered, h) and i), correspond to the wave function
renormalization. Their evaluation is straightforward, and their effect is taken
into account by multiplying the tree level contribution by $(1+\partial\Sigma
/\partial k^{0}$) (see Appendix B). Finally, diagrams j) and k) cancel at 
$\vec{q}=\vec{0}$ (see Appendix B), and therefore we do not consider them
here. Diagrams c) and d), containing the
vertex corrections, have been neglected in the evaluation of $G^{E}$ because
they are of order $1/M$ of the rest of the diagrams and terms of this order
(recoil corrections) have been omitted in the $\pi NN$, $\pi N\Delta$ and
$\pi\Delta\Delta$ vertices. However, we have evaluated numerically the
contribution of these terms, and although individually they are not too small we
find a very good cancellation between the terms involving a nucleon or a 
$\Delta$ propagator, thus justifying altogether the neglecting of all these
terms.

We should also point out that we have not included diagrams m) where the $\rho$
couples directly to the nucleon and there is a selfenergy insertion in the
$\rho$ corresponding to a two pion loop , or similarly diagrams l), where 
a tadpole of a pion loop is attached to the $\rho$ propagator. Such terms, as shown in 
\cite{Cabrera:2000dx,Oller:2000ug} go into the renormalization of the $\rho$
mass and width.

In table \ref{gegm} we summarize the results obtained here for $G^{E}$ and
$G^{M}$ at $\vec{q}=\vec{0}$.

\begin{table}
\begin{tabular}{|l|l|l|l|l|l|l|l|l|l|l|l|l|}
\hline
  &  tree & a) & b) & c) & d) & e) & f) & g) & h) & i) & total \\
\hline
  $G^{E}$ & 3.07 & 1.64 & -1.11 & -- & -- & -0.41 & -- & 2.77 & -1.23 & -1.66 &
  3.07 \\ 
  $G^{M}$ & 3.07 & 4.42 & 1.44 & -1.90 & -0.55 & 0.14 & 1.40 & 0.93 & -1.23 &
  -1.66 & 6.05 \\ 
\hline
\end{tabular}
\caption{Different contributions to $G^{E}$ and $G^{M}$ at $\vec{q}=\vec{0}$.}
\label{gegm}
\end{table}
 
It is worth noting that the different loop contributions approximately cancel with the
wave function renormalization (see table \ref{gegm}) in $G^{E}(\vec{q}=\vec{0})$.
The cancellation appears from diagrams a), e), h) as a block, which are the
terms containing intermediate nucleon propagators, and from diagrams b), g)
and i) which contain the $\Delta$ propagators. This cancellation can be found
analytically from the expressions in Appendix B at $\vec{q}=\vec{0}$ if we
include only one factor $M/E$ in the set of baryon propagators to keep at the
same order of a non-relativistic expansion.

The interesting thing to see in table \ref{gegm} is that the loops provide a
sizeable contribution to $G^{M}(\vec{q}=\vec{0})$ which has now a value of
6.05, about double the one from the tree level. This number is still small
compared to the empirical value of $G^{M}\sim 21$, and the discrepancy should be
attributed to non-perturbative effects.

In order to estimate the uncertainties we have changed
the value of the form factor parameter for the pion and the value of $D+F$. If we
use for instance $D+F=2ff_{\pi NN}/\mu$ instead of $D+F=g_{A}$ we obtain
$G^{M}(\vec{q}=\vec{0})=6.39$. If we change $\Lambda_{\pi}$ from 0.9 GeV to 1.2
GeV we obtain $G^{M}(\vec{q}=\vec{0})=6.41$. 
The results obtained by changing the parameters reasonably indicate that the 
uncertainties in the theoretical calculation are around as less than 10\%.

\begin{center}
\begin{figure}[h]
\psfig{file=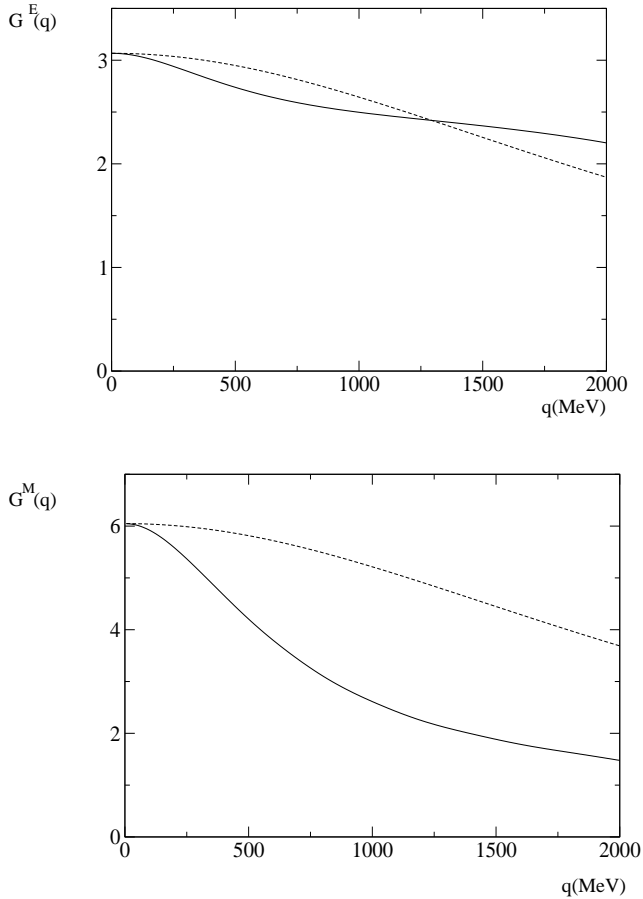,width=0.7\textwidth,silent=}
\label{ourff}
\caption{Comparison between the theoretical form factors (solid line) and the 
 empirical form factors as the one given in Appendix B (dashed
line).}
\end{figure} 
\end{center}

Finally, in figure \ref{ourff} we plot the results for the $\vec{q}$ dependence (form
factor) of $G^{E}(\vec{q})$ and $G^{M}(\vec{q})$. They are compared with the
empirical form factor, assuming the same normalization at $\vec{q}=\vec{0}$. For
the $\vec{q}$ dependence of our results we sum the tree level contribution
multiplied by the empirical form factor and the loop contributions evaluated
here. We can see that our calculation falls down faster than a monopole form
factor at low energies, perticularly the tensor part.

\section{Considerations on gauge invariance}

Although the procedure followed with the Feynman diagrams used would fulfill
gauge invariance, our introduction of the explicit form factors of Appendix B
would break it. This is a well known fact and in the literature there have been
many attempts to restore gauge invariance in the presence of form factors
\cite{Urban:1998eg,Berends:xw,Nacher:1998hh}.

Since our main concern is to show the contribution of loops to the magnetic
coupling (at $\vec{q}=\vec{0}$) or to $G^{E}$ and $G^{M}$ at moderate values of
$\vec{q}$, rather
than embarking in some of the procedures to restore gauge invariance as quoted
above, we shall make a study here of why and how gauge invariance is broken and
this will  give us an idea for which values of $\vec{q}$ our procedure still satisfies
gauge invariance and hence makes the results credible. Let us take the
contribution from all loop diagrams from a) to g) in fig.~\ref{loop} with an
incoming nucleon momentum $p$ and an outgoing nucleon momentum $p+q$, and call
their contribution to -i$t^{\mu}$:

\begin{center}
\be
-\textrm{i}\Delta t^{\mu}\equiv \textrm{i}\Delta G^{V}(q)\gamma^{\mu} + \frac{G^{T}(q)}{2
\textrm{i} M_{N}}\sigma^{\mu \nu}q_{\nu}
\label{unog}
\ee
\end{center}

A general test of gauge invariance can be given by the Ward identities which
establish in the $\rho NN$ case \cite{Urban:1998eg}

\begin{center}
\be
-\textrm{i}\Delta t^{\mu}(q)q_{\mu} =
\textrm{i}\frac{g_{\rho}}{2}\left( \Sigma(p+q)-\Sigma(p) \right)
\label{dosg}
\ee
\end{center}

\noindent where $\Sigma(p)$ is the nucleon self-energy. Eq.(\ref{dosg}) implies

\begin{center}
\be
\textrm{i}\Delta G^{V}(q)\gamma^{\mu}q_{\mu} = \textrm{i}\frac{g_{\rho}}{2} \left(\Sigma(p+q)-
\Sigma(p)\right)
\label{tresg}
\ee
\end{center}

\noindent which should be valid for any value of $p$ and $q$ even if the nucleons and the
$\rho$ are off shell. By chosing $p^{0}=M_{N}$, $\vec{p}=\vec{0}$ and arbitrary
$q^{0}$, $\vec{q}$, which is a sufficiently general situation, we have:

\begin{center}
\be
\textrm{i}\Delta G^{V}(q)\gamma^{\mu}q_{\mu} = \textrm{i}\frac{g_{\rho}}{2}
\left(\Sigma(q^{0}+M_{N},\vec{q})-\Sigma(M_{N},\vec{0})\right)
\label{cuatrog}
\ee
\end{center}

\noindent where in the non relativistic expansion which we are using
$\Sigma(p)$ is just a number. We now evaluate the matrix element of the first
member of eq.~(\ref{tresg}) between the spinors $\bar{u}(\vec{p}+\vec{q})$ and
$u(\vec{p})$ in the same non-relativistic approximation. Then we obtain

\begin{center}
\be
\textrm{i} \Delta G^{V}\left( q^{0}-\frac{\vec{q}^{\textrm{ }2}}{2M_N}\right) =
\textrm{i} \frac{g_{\rho}}{2} \left(\Sigma(q^{0}+M_{N},\vec{q})- 
\Sigma(M_{N},\vec{0})\right)
\label{cincog}
\ee
\end{center}

Let us now take $q^{0}$, $q$ small enough such that a Taylor expansion of 
$\Sigma$ can be done and we obtain

\begin{center}
\be
\textrm{i} \Delta G^{V}\left( q^{0}-\frac{\vec{q}^{\textrm{ }2}}{2M_N}\right) =
\textrm{i} \frac{g_{\rho}}{2} \left(q^{0}\left. \frac{\partial \Sigma}{\partial
q^{0}}\right|_{q^{0}=q=0} + \vec{q}^{\textrm{ }2}\left. \frac{\partial \Sigma}{\partial
\vec{q}^{\textrm{ }2}}\right|_{q^{0}=q=0}\right)
\label{seisg}
\ee
\end{center}

\noindent which by means of the easily derivable relation

\begin{center}
\be
\left. \frac{\partial{\Sigma}}{\partial \vec{q}^{\textrm{ }2}}\right|_{q^{0}=q=0} =
-\frac{1}{2M_{N}}\left. \frac{\partial \Sigma}{\partial q^{0}}\right|_{q^{0}=q=0}
\label{sieteg}
\ee
\end{center}

\noindent leads to

\begin{center}
\be
\textrm{i}\Delta G^{V}\left(q^{0}-\frac{\vec{q}^{\textrm{ }2}}{2M_N}\right) =
\textrm{i} \frac{g_{\rho}}{2} \left. \frac{\partial \Sigma}{\partial
q^{0}}\right|_{q^{0}=q=0}\left( q^{0}-\frac{\vec{q}^{\textrm{ }2}}{2M_N}\right)
\label{ochog}
\ee
\end{center}

This requires (omitting higher orders $q^{0}$, $\vec{q}$) that

\begin{center}
\be
\left. \Delta G^{V}(q)\right|_{q^{0}=q=0}-\frac{g_{\rho}}{2}\left.
\frac{\partial \Sigma}{\partial q^{0}}\right|_{q^{0}=q=0} = \left. 
\left(\Delta G^{E}+G^{E}_{tree}\frac{\partial \Sigma}{\partial q^{0}}\right)
\right|_{q^{0}=q=0}=0
\label{nueveg}
\ee
\end{center}

As we have noted in the former section, $\Delta
G^{E}+G^{E}_{tree}\frac{\partial\Sigma}{\partial q^{0}}$ gives the contribution
to $G^{E}(q)$ from the loop vertex functions of diagrams a) to g) of
fig.~\ref{loop},
plus the contribution from the wave function renormalization (diagrams h), i) of
fig.~\ref{loop}) which is given by the second term, 
$G^{E}_{tree}\frac{\partial\Sigma}{\partial q^{0}}$. Eq.~(\ref{nueveg}) tells us
that the contribution of all those terms to $G^{E}(0)$ is null, something that we
had already found numerically even with the presence of form factors. It is not
surprising that this should be the case because it should occur in the absence
of form factors from a cancellation of the contribution of the different
diagrams. Then at $q^{0}=q=0$ all these diagrams are multiplied by the same form
factors and hence the cancellation also holds. Since eq.~(\ref{nueveg}) is
satisfied also in the case of form factors, then eq.~(\ref{ochog}), which is the
statement of the Ward identities at moderate values of $q^{0}$ and $q$, also
holds in the presence of form factors. However, the limits go beyond those where
the Taylor expansion may hold. Indeed, as we stated above, the Ward identities
in the absence of form factors would be fulfilled and if all the diagrams were
multiplied in our case by the same form factors then the equality would still
hold. However, this is the case only at $q^{0}=q=0$, because for finite $q$ we
have in the loops $F_{\pi}^{2}(\vec{p})$ in some diagrams and
$F_{\pi}(\vec{p})F_{\pi}(\vec{p}+\vec{q})$ in other diagrams, which are not the same. Also
we have the $\rho NN$ form factor which does not appear in diagrams a) and b).
Since the form factor is only operative for values of $|\vec{p}|$ of the order of the
cut off $\Lambda$ or higher, demanding that the angle averaged value of 
$ F_{\pi}(\vec{p})F_{\pi}(\vec{p}+\vec{q})$ be similar to 
$F_{\pi}^{2}(\vec{p})$, for instance, implies that $q$ should be small compared
to the cut off scale $\Lambda$.

The former argument sets the scale of values of $q$ where gauge invariance would
be violated. On the other hand the results of \cite{Nacher:1998hh} using the
gauge restoring Berends formalism or the plain application of form factors
showed that the differences were not large even at values of $|\vec{q}|\sim 800$
MeV. All these things considered, one can reasonably say that up to values of 
$q\sim 500$ MeV the results evaluated here would be reliable.

\section{Conclusions}

We have evaluated the contributions to $G^{E}(\vec{q})$ and $G^{M}(\vec{q})$ for
the coupling of the $\rho$ meson to the nucleon including direct couplings
stemming from a gauge formulation of the theory and in addition we have included
the contributions of the virtual meson cloud at one loop level. We regularize
the loops by means of a form factor, introducing an effective cut off of the
order of 1 GeV which is considered the natural scale. In addition to this cut
off the space of the intermediate states is reduced to the nucleon and the
$\Delta$. This means a regularization from two sources and the justification
should be seen in the phenomenological success of such an approach in a large
number of hadronic properties in the evaluation of chiral bag models \cite{Thomas:1982kv}.

What we find, as expected, is that $G^{E}(\vec{q}=\vec{0})$ does not change with
respect to the tree level, because it is restricted by gauge conditions, but
$G^{M}(\vec{q}=\vec{0})$, which has no such restrictions, is appreciably
enhanced. However, this enhancement is still clearly insufficient to provide
values close to the empirical one. This result seems to indicate that the
magnetic coupling of the $\rho$ to the nucleon is of direct nature and cannot be
attributed to loop corrections. The $\rho$, as a genuine QCD state
\cite{Ecker:1989te}, by contrast to the low energy scalar mesons 
\cite{Oller:1997ti}, has also as a genuine property a strong magnetic coupling
to the nucleon, the origin of which goes beyond the meson loop calculation which
we have done. This is in contrast to the coupling of the $\sigma$ meson to the
nucleon which, in correspondence to the nature of the $\sigma$ as a pion-pion
rescattering resonance, could be obtained by coupling the meson cloud to the
nucleons \cite{Oset:2000gn,Jido:2001am}.

On the other hand we observe that the loop contribution to the
electric and magnetic form factors has a stronger $\vec{q}$ dependence than the
one provided by the assumed empirical monopole form factor. This is due to the
large extend of the pion cloud, which, due to the small mass of the pion, has a
larger range than the genuine constituents (quarks) of the $\rho$ meson. This
faster fall of the form factors, or, alternativelly the larger range of the
coupling, should also have some repercussion in the part of the $NN$ interaction
mediated by $\rho$ exchange.

\subsection*{Acknowledgments}

One of us, D.J. wishes to acknowledge the hospitality of the University of
Valencia where this work was done and financial support from the Ministerio de
Educacion in the program Doctores y Tecnologos extranjeros. J. E. P. wishes to
acknowledge support from Ministerio de Educaci\'on, Cultura y Deporte. 
This work is also partly
supported by DGICYT contract number BFM2000-1326 and E.U. EURODAPHNE network
contract no. ERBFMRX-CT98-0169.

\appendix
\section{$G^M_{\Delta}$ and $G^M_{N \Delta}$  from the
quark model\label{QM}} 

Here we explain the calculations of the $\rho \Delta \Delta$ and $\rho
N \Delta$ couplings from the $SU(6)$ quark model.
Now let us define the operator of the
$\rho^0$ coupling to the $i$-th quark for $G_M$: 
\begin{equation}
    \hat g_M^{(i)} = -\GMquark  
    {( \vec q \times \vec \varepsilon) \cdot \vec\sigma_{(i)}
    \over 2 m_q} \tau^3_{(i)}   
\end{equation}
with the light quark mass $m_q$ and an outgoing $\rho^0$ momentum
$\vec q$. The $G^M$ for proton with up spin is calculated from the
quark model as 
\begin{equation}
   \langle p \uparrow | \sum_{i=1}^3 \hat g_M^{(i)} | p \uparrow
   \rangle = -{5 \over 3} \GMquark  { (\vec q \times \vec
   \varepsilon)_3 \over 2 m_q}
\end{equation}
Here we use $\langle p \uparrow |\sigma^3 \tau^3| p \uparrow
\rangle={5\over 3}$ and  the $SU(6)$ flavor-spin symmetry for the
nucleon wave function:   
\begin{equation}
    | N \rangle = {1 \over \sqrt{2}}(\phi_{MS} \chi_{MS} + \phi_{MA}
      \chi_{MA}) 
\end{equation}
with, for the proton with the spin up, 
\begin{eqnarray}
   \phi_{MS}^{(p)} = {1 \over \sqrt{6}} [ (ud+du)u-2uud] &\hspace{0.5cm}&
   \phi_{MA}^{(p)} = {1 \over \sqrt{2}} (ud-du)u\\
   \chi_{MS}^{(\uparrow)} = {1 \over \sqrt{6}} [(\uparrow\downarrow +
   \downarrow\uparrow) \uparrow -2 \uparrow\uparrow\downarrow] &\hspace{0.5cm}&
   \chi_{MA}^{(\uparrow)} = {1 \over \sqrt{2}}
   (\uparrow\downarrow-\downarrow\uparrow)\uparrow 
\end{eqnarray}
Comparing with the definition of the magnetic coupling for nucleon
(\ref{GEGM}), we obtain the relation of the $G^M_{N}$ to the
$\GMquark$: 
\begin{equation}
   {\GMquark \over 2 m_q} = {3 \over 5} {G^M_N \over 2 \Nmass}
\end{equation}

In the same way, the $\rho N\Delta$ and $\rho \Delta \Delta$ couplings
are calculated with the quark model. Using $ \langle p \uparrow
|\sigma^3 \tau^3 | \Delta^+\, {\textstyle{1\over 2}}
\rangle={8 \over 3 \sqrt{2}}$ and $ \langle \Delta^+\,
{\textstyle{1\over 2}}|\sigma^3\tau^3 | \Delta^+\, {\textstyle{1\over 2}}
\rangle ={1\over3} $, we obtain
\begin{eqnarray}
   \langle  p \uparrow |\sum_{i=1}^3 \hat g_M^{(i)}|\Delta^+\,
   {\textstyle{1\over 2}}\rangle &= & -{8 \over 3 \sqrt{2}} \GMquark 
   { (\vec q \times \vec \varepsilon)_3 \over 2 m_q} \\
   \langle \Delta^+\,{\textstyle{1\over 2}}|\sum_{i=1}^3 \hat
   g_M^{(i)}| \Delta^+\, {\textstyle{1\over 2}}\rangle &=& -{1\over3}
   \GMquark  { (\vec q \times \vec \varepsilon)_3 \over 2 m_q}
\end{eqnarray}
Here the $SU(6)$ wave function for the $\Delta^+$ with spin $1/2$ is
given by
\begin{equation}
   | \Delta^+\, {\textstyle{1\over 2}}
\rangle ={1 \over \sqrt{3}}(uud+udu+duu) {1 \over
\sqrt{3}}(\uparrow\uparrow\downarrow + \uparrow\downarrow\uparrow +
\downarrow\uparrow\uparrow) 
\end{equation} 
Taking care of the normalization of the definitions of the couplings
(\ref{rhoDD}) and (\ref{rhoDN}), where $  \langle p \uparrow
|-S^3 T^3 | \Delta^+\, {\textstyle{1\over 2}} \rangle=-\sqrt{2\over 3}
\sqrt{2\over 3}$  and $ \langle \Delta^+\, {\textstyle{1\over 2}} |-
S^3_\Delta T^3_\Delta | \Delta^+\, {\textstyle{1\over 2}} \rangle =
-{1\over 2} {1 \over 2} $, finally we obtain the relations to the $\rho
NN$ coupling:
\begin{equation}
   G^M_{N \Delta} = {6 \over 5 } \sqrt{2} G^M_{N} 
   \hspace{0.7cm}
   G^M_{ \Delta} = {4 \over 5 } G^M_{N}
\end{equation}

\section{ One loop calculations}

In this Appendix we give the explicit expressions of the contributions of the
loop diagrams to $G^{V}_{NN\rho}$ and $G^{M}_{NN\rho}$. In the following
equations and diagrams 
$\epsilon_{\mu}$ denotes the $\rho$ polarization vector, and:

\begin{center}
\ba
q\equiv (E(\vec{q}),\vec{q}) \textrm{\ \ \ \ \ \ \ } & & Q\equiv (0,\vec{q}) \nn \\
  \omega(k)\equiv\sqrt{\vec{k}^{2}+m_{\pi}^{2}} \textrm{\ \ \ \ \ }& &
  D(k)\equiv\frac{1}{k^{2}-m_{\pi}^{2}} \nn \\
F_{\pi}(\vec{k})\equiv \frac{\Lambda^{2}}{\Lambda^{2}+\vec{k}^{2}} \textrm{ \ \ \ \ }
& &F_{\rho}(\vec{k})\equiv \frac{\Lambda_{\rho}^{2}}
{\Lambda_{\rho}^{2}+\vec{k}^{2}}\nn \\
\Lambda=0.9\textrm{ GeV \ \ \ \ \ \ \ \ }  & &\Lambda_{\rho}=2.5\textrm{ GeV}\nn \\
E(\vec{k})\equiv\frac{\vec{k}^{\textrm{ }2}}{2M_{N}}+M_{N}\textrm{\ \ \ \ \ \ \ \ \ } & & 
E_{\Delta}(\vec{k})\equiv \frac{\vec{k}^{\textrm{ }2}}{2M_{\Delta}}+M_{\Delta}
\label{defi}
\ea
\end{center}

We warn the reader that, in order not to complicate excessively the expressions,
we have deliberately omitted the form factors and the $M/E$ relativistic
corrections to the baryonic propagators in the following equations,
although it should be kept in mind that one must include them to perform the
numerical calculations.

   \begin{figure}[h]
   \psfig{file=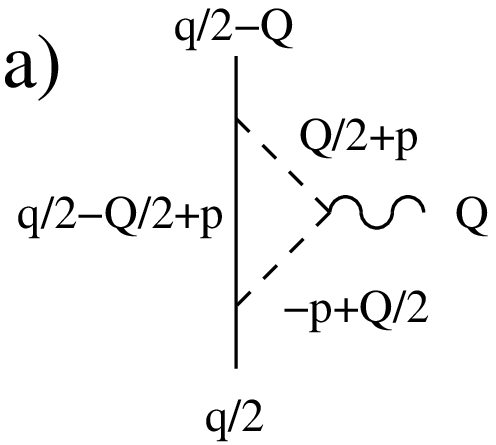,width=0.25\textwidth,silent=}
   \label{geNfig}
   \end{figure}    

\ba
-iV(\vec{q})=-4\epsilon_{\mu}\left(\frac{g_{A}}{2f}\right)^{2}
g_{\rho}\int\frac{d^{4}p}{(2\pi)^{4}}\frac{\vec{\sigma}
(\vec{p}-\vec{q}/2)\vec{\sigma}(\vec{p}+\vec{q}/2)}{E(\vec{q}/2)+p^{0}
-E(\vec{p})+i\epsilon}\times&\nn \\ \times
D(p-q/2)D(p+q/2)p^{\mu}
\label{va}
\ea

\vspace{1cm}



\begin{center}
\be
G^{E\textrm{ }a)}_{\rho NN}(\vec{q})=-2 \left(\frac{g_{A}}{2f}\right)^{2} g_{\rho}
\int\frac{d^{3}p}{(2\pi)^{3}}\left(\vec{p}^{\textrm{ }2}-\frac{\vec{q}^{\textrm{ }2}}{4}\right)
f_{1}(\vec{p},\vec{q})
\label{geN}
\ee
\end{center}

\begin{center}
\be
\frac{G^{M \textrm{ }a)}_{\rho NN}}{2M_{N}}=-2\left(\frac{g_{A}}{2f}\right)^{2} g_{\rho} 
\int\frac{d^{3}p}{(2\pi)^{3}}\left(\vec{p}^{\textrm{ }2}-\frac{(\vec{p}
\vec{q})^{2}}{\vec{q}^{\textrm{ }2}}\right) f_{2}(\vec{p},\vec{q})
\label{gmN}
\ee
\end{center}

\noindent where the $f_1$ and $f_2$ functions are defined as:

\begin{center}
\ba
f_{1}(\vec{p},\vec{q})&=&\frac{1}{\omega(\vec{p}+\vec{q}/2)+\omega(\vec{p}-
\vec{q}/2)}\textrm{ }\frac{1}{E(\vec{q}/2)-\omega(\vec{p}+\vec{q}/2)-E(\vec{p})}
\times \label{efes} \\ & \times&\frac{1}{E(\vec{q}/2)-\omega(\vec{p}-\vec{q}/2)-E(\vec{p})}\nn \\
f_{2}(\vec{p},\vec{q})&=&\frac{1}{\omega(\vec{p}+\vec{q}/2)+\omega(\vec{p}-
\vec{q}/2)}\textrm{ }\frac{1}{E(\vec{q}/2)-\omega(\vec{p}+\vec{q}/2)-E(\vec{p})}
\times \nn \\ &\times &\frac{\omega
(\vec{p}+\vec{q}/2)+\omega(\vec{p}-\vec{q}/2)+E(\vec{p})-E(\vec{q}/2)}
{E(\vec{q}/2)-\omega(\vec{p}-\vec{q}/2)-E(\vec{p})}
 \textrm{ }\frac{1}{2\omega(\vec{p}+\vec{q}/2)\omega(\vec{p}-\vec{q}/2)}
\nn
\ea
\end{center}
\vspace{0.3cm}

\begin{figure}[h]
\psfig{file=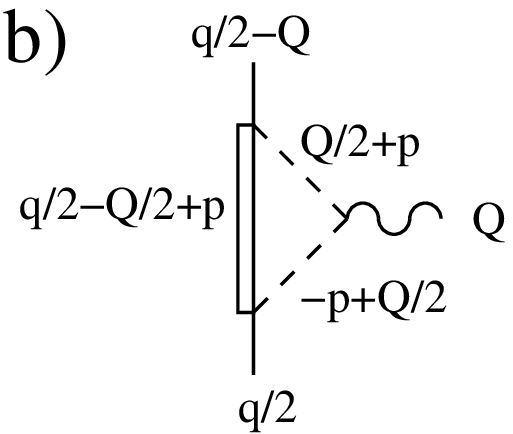,width=0.25\textwidth,silent=}
\label{gedelfig}
\end{figure} 

In the calculation of diagrams with intermediate $\Delta$'s one has different spin
and isospin factors since the spin and isospin transition operators appearing in
the corresponding Lagrangians satisfy the following relations:

\begin{center}
\begin{eqnarray}
S_{i}S^{\dagger}_{j}=\frac{2}{3}\delta_{ij}-
\frac{i}{3}\epsilon_{ijk}\sigma_{k}\nn \\
T_{i}T^{\dagger}_{j}=\frac{2}{3}\delta_{ij}-
\frac{i}{3}\epsilon_{ijk}\tau_{k}
\label{repchange}
\ea
\end{center}

Taking this into account one finds

\begin{center}
\be
G^{E\textrm{ }b)}_{\rho NN}(\vec{q})=\frac{4}{9}\left( \frac{f^{*}_{\pi N\Delta}}{f_{\pi
NN}}\right)^{2} \left(\frac{g_{A}}{2f}\right)^{2} g_{\rho}
\int\frac{d^{3}p}{(2\pi)^{3}}\left(\vec{p}^{\textrm{ }2}-\frac{\vec{q}^{\textrm{
}2}}{4}\right)
f_{1}^{\Delta}(\vec{p},\vec{q})
\label{gedel}
\ee
\end{center}

\begin{center}
\be
\frac{G^{M\textrm{ }b)}_{\rho NN}}{2M_{N}}=-\frac{2}{9}\left( \frac{f^{*}_{\pi N\Delta}}{f_{\pi
NN}}\right)^{2}\left(\frac{g_{A}}{2f}\right)^{2} g_{\rho} 
\int\frac{d^{3}p}{(2\pi)^{3}}\left(\vec{p}^{\textrm{ }2}-\frac{(\vec{p}
\vec{q})^{2}}{\vec{q}^{\textrm{ }2}}\right) f_{2}^{\Delta}(\vec{p},\vec{q})
\label{gmdel}
\ee
\end{center}

\noindent where $f_{1}^{\Delta}$ and $f_{2}^{\Delta}$ are defined as $f_{1}$ and
$f_{2}$ (see eqs. (\ref{efes})), but replacing there $E(\vec{p})$ by
$E_{\Delta}(\vec{p})$, and $f^{*}_{\pi N\Delta}/f_{\pi N}=2.12$.
\vspace{0.6cm}

\hspace{-1cm}\parbox{3cm}{
\psfig{file=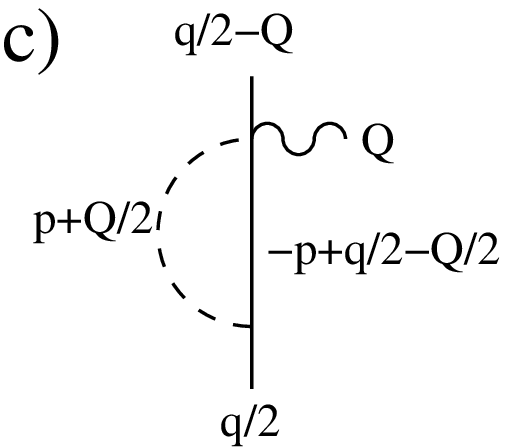,width=0.25\textwidth,silent=}

\label{gepefig}
}
\parbox{10cm}{
\be
G^{E\textrm{ }c)}_{\rho NN}(\vec{q})={\cal O}(1/M_{N})
\ee}

\begin{center}
\ba
\textrm{ \ }\frac{G^{M\textrm{ }c)}_{\rho NN}}{2M_{N}}&=&-2\left(\frac{g_{A}}{2f}\right)^{2} g_{\rho}
\int\frac{d^{3}p}{(2\pi)^{3}}\left(1+\frac{2\vec{p}\vec{q}}{\vec{q}^{2}}\right)\frac{1}{2\omega
(\vec{p}+\vec{q}/2)}\times \nn \\ &\times & \frac{1}{E(\vec{q}/2)-
\omega(\vec{p}+\vec{q}/2)-E(\vec{p})}
\label{gmpe}
\ea
\end{center}
\vspace{0.6cm}

\hspace{-1cm}\parbox{3cm}{
\psfig{file=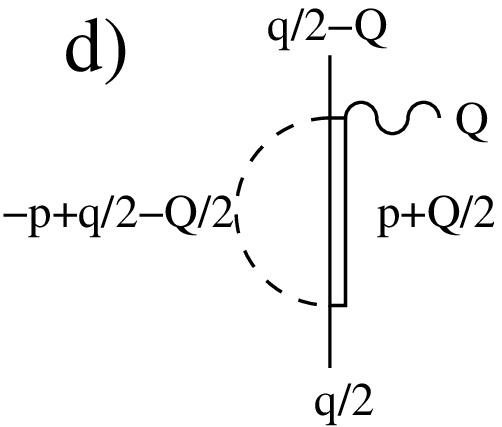,width=0.25\textwidth,silent=}
\label{geoperupfig}
}
\parbox{10cm}{\be
G^{E\textrm{ }d)}_{\rho NN}(\vec{q})={\cal O}(1/M_{N})
\ee}

\begin{center}
\ba
\textrm{ \ \ }\frac{G^{M\textrm{ }d)}_{\rho NN}}{2M_{N}}&=&-\frac{2}{9}\left( \frac{f^{*}_{\pi N\Delta}}{f_{\pi
NN}}\right)^{2}\left(\frac{g_{A}}{2f}\right)^{2} g_{\rho}
\int\frac{d^{3}p}{(2\pi)^{3}}\left(1+\frac{2\vec{p}\vec{q}}
{\vec{q}^{2}}\right)
\times  \nn \\ &\times& \frac{1}{2\omega(\vec{p}+\vec{q}/2)} 
\frac{1}{E(\vec{q}/2)-
\omega(\vec{p}+\vec{q}/2)-E_{\Delta}(\vec{p})} \label{gmoperup}
\ea
\end{center}
\vspace{0.6cm}

\hspace{-1cm}\parbox{3cm}{
\psfig{file=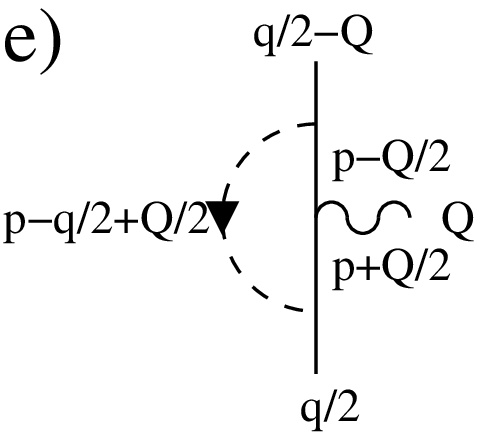,width=0.25\textwidth,silent=}
\label{ge1fig}
}
\parbox{10cm}{
\ba
\textrm{ \ \ }G^{E\textrm{ }e)}_{\rho NN}(\vec{q})=\frac{1}{2}\left(\frac{g_{A}}{2f}\right)^{2}g_{\rho}
\int\frac{d^{3}p}{(2\pi)^{3}}\frac{1}{2\omega(\vec{p})} 
\times \nn \\ \times \frac{\vec{p}^{\textrm{ }2}}{E(\vec{q}/2)-\omega(\vec{p})-E(\vec{q}/2-\vec{p})}
\frac{1}{ E(\vec{q}/2)-\omega(\vec{p})-E(-\vec{q}/2-\vec{p})}
\label{ge1}
\ea
}
\begin{center}
\ba
G^{M\textrm{ }e)}_{\rho NN}(\vec{q})&=&-\frac{g_{\rho}}{2} 
\left(\frac{g_{A}}{2f}\right)^{2}\int\frac{d^{3}p}
{(2\pi)^{3}}\frac{(\vec{p}\vec{q})^{2}}{2\vec{q}^{\textrm{ }2}\omega(\vec{p})}
 \frac{1}{E(\vec{q}/2)-\omega(\vec{p})-E(\vec{q}/2-\vec{p})}
\times \nn \\ &\times&\frac{1}{ E(\vec{q}/2)-\omega(\vec{p})-E(-\vec{q}/2-\vec{p})}
\label{gm1}
\ea
\end{center}

\hspace{-1cm}\parbox{3cm}{
\psfig{file=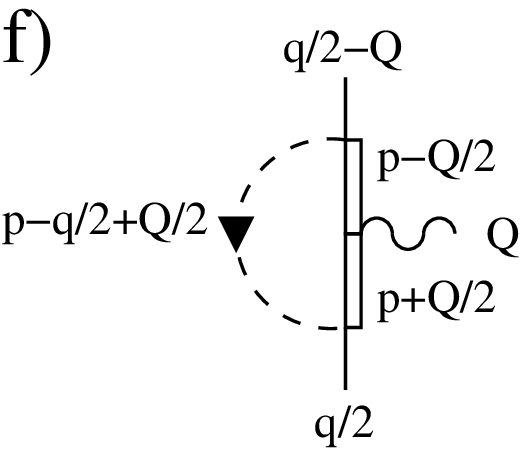,width=0.25\textwidth,silent=}
\label{ge1dfig}
}
\parbox{10cm}{\ba
\textrm{\hspace{-0.3cm}}G^{E\textrm{ }f)}_{\rho NN}(\vec{q})=-\frac{10}{9}\left(\frac{g_{A}}{2f}\right)^{2}g_{\rho}
\left( \frac{f^{*}_{\pi N\Delta}}{f_{\pi
NN}}\right)^{2}\int\frac{d^{3}p}{(2\pi)^{3}}\frac{1}{2\omega(\vec{p})} \times 
 \label{ge1d} \\ \times 
\frac{\vec{p}^{\textrm{ }2}}{E(\vec{q}/2)-\omega(\vec{p})-E_{\Delta}(\vec{q}/2-
\vec{p})}\textrm{ }
\frac{1}{ E(\vec{q}/2)-\omega(\vec{p})-E_{\Delta}(-\vec{q}/2-\vec{p})}
\nn
\ea}

\begin{center}
\ba
G^{M\textrm{ }f)}_{\rho NN}(\vec{q})&=&-\frac{1}{3}g_{\rho}\left(\frac{g_{A}}{2f}\right)^{2}\left( 
\frac{f^{*}_{\pi N\Delta}}{f_{\pi NN}}\right)^{2}\int\frac{d^{3}p}{(2\pi)^{3}}
\left( \vec{p}^{\textrm{ }2}+\frac{(\vec{p}\vec{q})^{2}}{3\vec{q}^{\textrm{ }2}}
\right) \frac{1}{2\omega(\vec{p})} \times  \nn \\
&\times &\frac{1}{E(\vec{q}/2)-\omega(\vec{p})-E_{\Delta}(\vec{q}/2-\vec{p})}
\textrm{ }
\frac{1}{E(\vec{q}/2)-\omega(\vec{p})-E_{\Delta}(-\vec{q}/2-\vec{p})}
\nn \\ \label{gm1d}
\ea
\end{center}
\vspace{0.3cm}

\hspace{-1cm}\parbox{3cm}{
\psfig{file=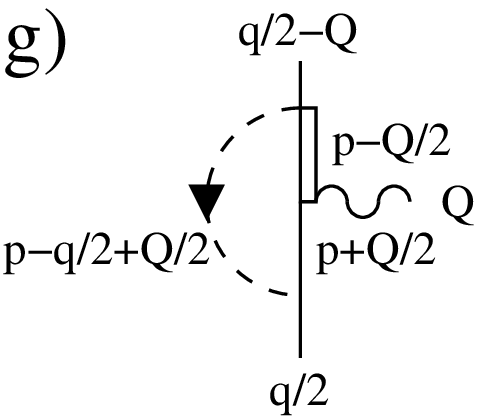,width=0.25\textwidth,silent=}
\label{ge1Ndfig}
}
\parbox{10cm}{\be
G^{E\textrm{ }g)}_{\rho NN}(\vec{q})=0
\ee}

\begin{center}
\ba
G^{M\textrm{ }g)}_{\rho NN}(\vec{q})&=&-\frac{8\sqrt{2}}{5}\frac{g_{\rho}}{2}
\left(\frac{g_{A}}{2f}\right)^{2}\left( 
\frac{f^{*}_{\pi N\Delta}}{f_{\pi NN}}\right)\int\frac{d^{3}p}{(2\pi)^{3}}
\left( \frac{\vec{p}^{\textrm{ }2}}{2}-\frac{(\vec{p}\vec{q})^{2}}{6\vec{q}
^{\textrm{ }2}}
\right) \frac{1}{2\omega(\vec{p})} \times \nn \\&\times &
\left\{\frac{1}{E(\vec{q}/2)-\omega(\vec{p})-E(\vec{q}/2-\vec{p})}
\textrm{ }\frac{1}{E(\vec{q}/2)-\omega(\vec{p})-E_{\Delta}(-\vec{q}/2-\vec{p})}
+ \right. \nn \\ &+& \left.
\frac{1}{E(\vec{q}/2)-\omega(\vec{p})-E_{\Delta}(-\vec{q}/2-\vec{p})}\textrm{ }
\frac{1}{E(\vec{q}/2)-\omega(\vec{p})-E(\vec{q}/2-\vec{p})} \right\}\nn \\
\label{gm1Nd}
\ea
\end{center}
\vspace{0.3cm}

\hspace{-1cm}\parbox{3cm}{
\psfig{file=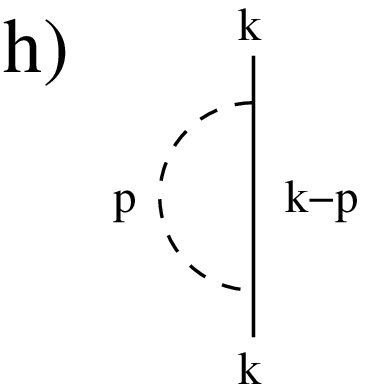,width=0.20\textwidth,silent=}
\label{zetafig}
}
\begin{center}
\be
\left. \frac{\partial \Sigma^{N}_{N}}{\partial k^{0}}\right|_{k^{0}=M_{N}}=
-3\left(\frac{g_{A}}{2f}\right)^{2}\int\frac{d^{3}p}{(2\pi)^{3}}\frac{\vec{p}^{\textrm{
}2}}{2\omega(\vec{p})[M_{N}-\omega(\vec{p})-E(\vec{k}-\vec{p})]^{2}}
\label{zeta}
\ee
\end{center}

\vspace{0.3cm}

\begin{figure}[h]
\psfig{file=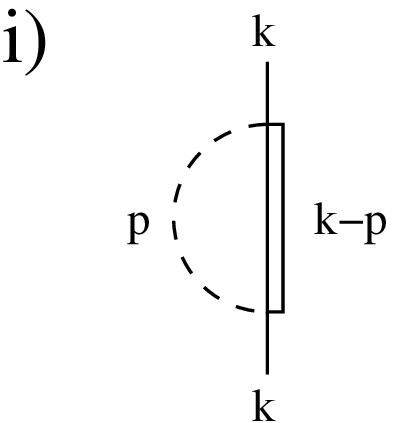,width=0.20\textwidth,silent=}
\label{zetadelfig}
\end{figure} 

\begin{center}
\be
\left. \frac{\partial \Sigma^{\Delta}_{N}}{\partial k^{0}}\right|_{k^{0}=M_{N}}=
-\frac{4}{3}\left(\frac{g_{A}}{2f}\right)^{2}\left( 
\frac{f^{*}_{\pi N\Delta}}{f_{\pi NN}}\right)^{2}
\int\frac{d^{3}p}{(2\pi)^{3}}\frac{\vec{p}^{\textrm{
}2}}{2\omega(\vec{p})[M_{N}-\omega(\vec{p})-E_{\Delta}(\vec{k}-\vec{p})]^{2}}
\label{zetadel}
\ee
\end{center}

\vspace{0.3cm}
\begin{figure}[h]
\psfig{file=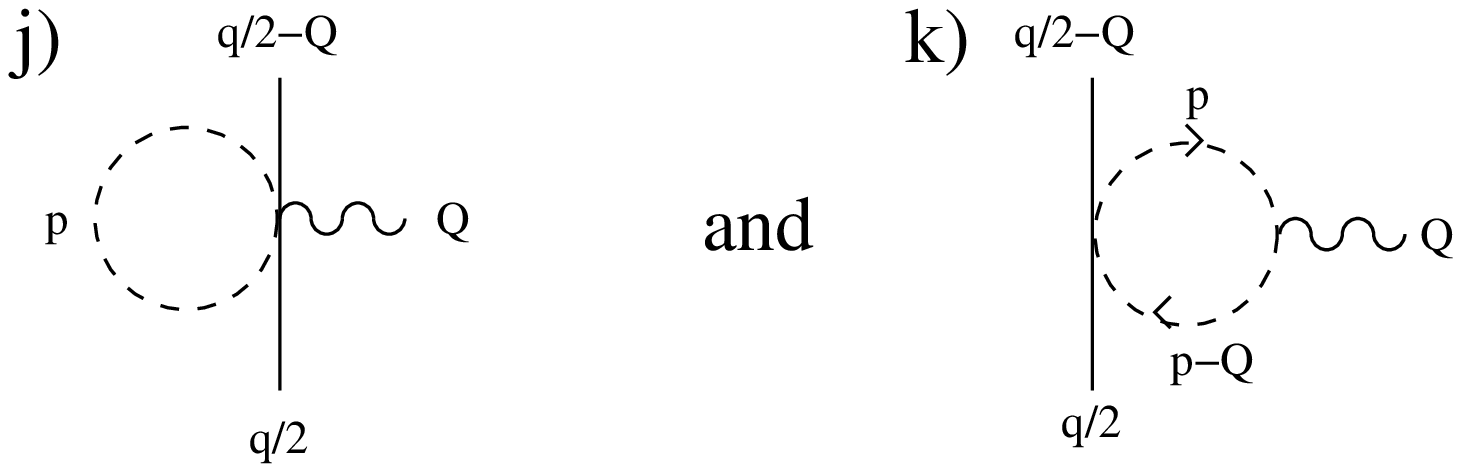,width=0.85\textwidth,silent=}
\label{cance}
\end{figure} 

We do not take into account these two diagrams since they cancel at
$\vec{q}=\vec{0}$. At this value of $\vec{q}$ diagram j) is proportional to:

\begin{center}
\begin{equation}
\int\frac{d^{4}p}{(2\pi)^{4}}2\gamma^{\mu}g_{\mu\nu}\epsilon^{\nu }D(p)
\end{equation}
\end{center}

\noindent and diagram k) is proportional to:

\begin{center}
\begin{equation}
-\int\frac{d^{4}p}{(2\pi)^{4}}\gamma^{\mu}4p_{\mu}\epsilon^{\nu }p_{\nu}D(p)D(p)
\label{jyk}
\end{equation}
\end{center}

Taking into account the integral identity:

\begin{center}
\be
\int {d^{4}p}\frac{4p^{\mu}p^{\nu}}{(p^{2}+s+i\epsilon)^{2}} = \int
d^{4}p\frac{2g^{\mu\nu}}{k^{2}+s+i\epsilon}
\label{relation}
\ee
\end{center}

\noindent it is straightforward to see that these diagrams cancel at
$\vec{q}=\vec{0}$.

\end{document}